\documentclass[letterpaper, 10 pt, conference]{ieeeconf}
\IEEEoverridecommandlockouts
\usepackage[utf8]{inputenc}
\usepackage{mathtools}
\usepackage{amsfonts}
\usepackage{xcolor}
\let\labelindent\relax % solve conflict between IEEE and enumitems
\usepackage{enumitem}
\usepackage{cite}
\usepackage{soul}
\usepackage{dsfont}
\usepackage{subfigure}
\usepackage{color}
\usepackage{float}

\usepackage{graphicx}
\graphicspath{ {./figures/} }

\usepackage{hyperref}
\hypersetup{
    colorlinks=true,
    linkcolor=blue,
    filecolor=magenta,      
    urlcolor=cyan,
}

\title{\LARGE \bf
Traffic Smoothing Controllers for Autonomous Vehicles Using Deep Reinforcement Learning and Real-World Trajectory Data
}

\author{Nathan Lichtlé$^{1,2\dagger*}$, Kathy Jang$^{1\dagger}$, Adit Shah$^{1\dagger}$, Eugene Vinitsky$^{3\dagger}$, \\ Jonathan W.~Lee$^{1,4}$, and Alexandre M.~Bayen$^{1,4}$%
\thanks{This work was supported in part by the C3.ai Digital Transformation Institute under Grant Number 053483. Nathan Lichtlé is supported in part by the International Emerging Actions project SHYSTRA (CNRS). This material is based upon work supported by the National Science Foundation under Grants CNS-1837244 (K.~Jang), CNS-2146752 (A.~Bayen), CNS-2135579 (A.~Bayen, J.~Lee). This material is based upon work supported by the U.S.\ Department of Energy’s Office of Energy Efficiency and Renewable Energy (EERE) under the Vehicle Technologies Office award number CID DE--EE0008872. The views expressed herein do not necessarily represent the views of the U.S.\ Department of Energy or the United States Government.}%
\thanks{}%
\thanks{$^{1}$ Department of Electrical Engineering and Computer Sciences, UC Berkeley}%
\thanks{$^{2}$ CERMICS, École des Ponts ParisTech}%
\thanks{$^{3}$ Department of Mechanical Engineering, UC Berkeley}%
\thanks{$^{4}$ Institute for Transportation Studies, UC Berkeley}%
\thanks{}%
\thanks{$^{\dagger}$ \textbf{These authors contributed equally to this work.}}%
\thanks{$^*$ Corresponding author: {\tt\small nathanlct@berkeley.edu}}%
}

\begin{document}

\maketitle
\thispagestyle{empty}
\pagestyle{empty}

\begin{abstract}
Designing traffic-smoothing cruise controllers that can be deployed onto autonomous vehicles is a key step towards improving traffic flow, reducing congestion, and enhancing fuel efficiency in mixed autonomy traffic.
We bypass the common issue of having to carefully fine-tune a large traffic micro-simulator by leveraging real-world trajectory data from the I-24 highway in Tennessee, replayed in a one-lane simulation. 
Using standard deep reinforcement learning methods, we train energy-reducing wave-smoothing policies. As an input to the agent, we observe the speed and distance of only the vehicle in front, which are local states readily available on most recent vehicles, as well as non-local observations about the downstream state of the traffic. We show that at a low 4\% autonomous vehicle penetration rate, we achieve significant fuel savings of over 15\% on trajectories exhibiting many stop-and-go waves. Finally, we analyze the smoothing effect of the controllers and demonstrate robustness to adding lane-changing into the simulation as well as the removal of downstream information.

\end{abstract}

\section{INTRODUCTION}

Transportation accounts for a large share of energy usage worldwide, with the U.S. alone attributing 28\% of its total energy consumption in 2021 to moving people and goods~\cite{eiaEnergyTransportation}. It is the single largest sector of energy consumption, ranking over other major contributors such as the industrial sector~\cite{eiaUSEnergy}. 
Advances in technology have paved the way for improvements in $CO_2$ emissions and fuel economy via consumer shifts toward hybrid and electric vehicles (EVs), as well as innovation in vehicle technology such as turbocharged engines or cylinder deactivation~\cite{epaHighlightsAutomotive}. Meanwhile, as autonomous vehicles (AVs) become increasingly more available in roadways, with the Insurance Institute for Highway Safety predicting there to be 3.5 million vehicles with autonomous capabilities on U.S. roads by 2025~\cite{naicAutonomousVehicles}, so too do their potential to yield a pronounced effect on the state of traffic, addressing the problem of energy usage from another angle.

This paper focuses on the problem of energy usage in transportation and explores the potential of AVs to alleviate this issue. In this era of mixed autonomy traffic, in which a percentage of vehicles are AVs with special control capabilities, a rich amount of work has been produced that shows that even a small percentage of intelligent agents in traffic are capable of achieving energy savings and traffic reduction via wave dampening or jam-absorption driving of stop-and-go waves and bottlenecks~\cite{beaty1998traffic, wu2019tracking, Jang2018SimulationVehicles, vinitsky2018lagrangian}. Many studies have shown how longitudinal control of an AV can significantly impact global and local metrics such as total velocity or fuel economy; for instance, controlled vehicles are capable of completely dissipating traffic waves in a ring setting~\cite{stern2018dissipation, cui2017stabilizing}. To our knowledge, there have no been no large-scale tests conducted with the goal of smoothing traffic flow.  

However, finding an effective way to design controllers for these settings remains an open question due to the partially observed, hybrid, multi-agent nature of traffic flow.
In recent years, reinforcement learning (RL) has emerged as a powerful approach for traffic control, leveraging its ability to capture patterns from unstructured data. RL is responsible for producing quality controllers across a wide range of domains, from robotics~\cite{gu2017deep} to mastering gameplay over human experts such as with Starcraft or Go~\cite{vinyals2019grandmaster, silver2017mastering}. Within the domain of traffic, RL has been used to derive a variety of state-of-the-art controllers for improving traffic flow metrics such as throughput and energy efficiency~\cite{wu2017flow, vinitsky2018benchmarks, Jang2018SimulationVehicles}.

With a particular focus on the real-world impacts of RL on traffic, we discuss the RL-based controller we developed for a stretch of the I-24 highway near Nashville, Tennessee. 
Our controllers take into account the requirements of real-world deployment, utilizing observations that are accessible via radar and cameras.
Despite limited access to the full state space, our RL-based approach achieves notable fuel savings even with low penetration rates.

The contributions of this article are:
\begin{itemize}
    \item The introduction of a single-agent RL-based controller developed using real traffic trajectories with advanced telemetry-enabled downstream information,
    \item Numerical results of the controller's performance demonstrating significant fuel savings of over 15\% in scenarios exhibiting large-amplitude stop-and-go waves.
\end{itemize}

\section{PROBLEM FORMULATION}

In this paper, we consider the problem of decreasing energy consumption in highway traffic by attempting to smooth out stop-and-go waves. These waves are a phenomenon where high vehicle density causes vehicles to start and stop intermittently, creating a wave-like pattern that can propagate upstream and be highly energy-inefficient due to frequent accelerating and braking~\cite{sugiyama2008traffic}. We insert a small percentage of autonomous vehicles (AVs) equipped with reinforcement learning-based controllers into the flux of traffic. Leveraging a data-driven, one-lane simulator previously introduced in~\cite{lichtle2022deploying}, we simulate real-world highway trajectories. This approach is considerably more time-efficient than comprehensive traffic micro-simulations and is able to partially model the intricate stop-and-go behaviors that occur in traffic, although it overlooks complex dynamics such as lane-changing, which we rectify by incorporating a lane-changing model that we calibrate on data. The following subsections introduce the simulation as well as the different modules that it integrates.

\subsection{Dataset and Simulation}

We use the I-24 Trajectory Dataset~\cite{matthew_nice_2021_6456348} introduced in~\cite{lichtle2022deploying} along with a one-lane simulator that replays collected trajectory data. A vehicle, which we call the \emph{trajectory leader}, is placed at the front of the platoon and replays the real-world I-24 trajectory. Behind it, we place an arbitrary number of AVs and human vehicles (introduced in Sec.~\ref{sec:idm}). Typically during training, the platoon behind the trajectory leader consists of one RL-controlled AV, followed by 24 human vehicles. The goal of the AV is to absorb the perturbations in the leader trajectory, so that the energy consumption of the following human vehicles improves compared to the case where the AV is not present. 

Instead of training on the whole dataset, we only train on a selected set of four trajectories containing different patterns of waves alternating between low and high speeds. This allows us to optimize directly on the dynamics we are interested in improving, without having training diluted by free-flow trajectories containing no waves for the controller to smooth. We observed that this made training faster while still yielding a controller able to generalize to unseen trajectories. As the trajectories are quite long (between 5000 and 12000 time steps, where a time step is $0.1$s), each simulation randomly samples a chunk of size 500 (or 50s) within a trajectory. At evaluation time, we consider a set of six trajectories distinct from the training trajectories and simulate the whole trajectories.

\subsection{Speed planner}

We use real-time data about the state of traffic all along the I-24 in order to equip the RL control with some knowledge about the downstream state of traffic. The data is provided to us by INRIX, which broadcasts it in real time with an approximately 1-minute update interval and a 3-minute delay. The highway is divided into segments of 0.5 miles on average (with significant variance), and the data consists of the average speed of traffic in each segment. For training, we retrieved the INRIX data matching the time when the trajectories were collected on the highway (which includes delay). In the absence of such historical data, one could also generate synthetic traffic data for each dataset trajectory by artificially and averaging the trajectory speeds accordingly.

On top of this raw INRIX data, we use a \emph{speed planner} developed in~\cite{fu2023planner} that provides a profile that the controller should approximately track. The speed planner takes in the data and interpolates on the individual data points to create a continuous and less noisy speed profile of the whole highway. It then uses kernel smoothing to create a \emph{target speed profile}, which is intended as an estimate of the speed to drive at in order to smooth out the traffic waves. However, due to delay and noisy estimates, driving exactly at this speed is insufficient to guarantee smoothing or reasonable gaps. This target speed profile, sampled at different points, is finally fed as an input to our controller, which can be used as an indication of where the free-flow and the congested regions might be.

\subsection{Energy function}

Since we aim to optimize energy consumption, we need a model that can be used to compute consumed energy in our cost functions. We use a model of instantaneous fuel consumption, a polynomial fitted on a Toyota RAV4 model that is similar to the model in~\cite{lee2021energy}, but with updated coefficients.
The function depends on the speed of the AV and its instantaneous acceleration, as well as road grade, which we assume to be zero in this work. 

\subsection{Human Driver Behavior}
\label{sec:idm}

To model human drivers in simulation, we use the Intelligent Driver Model (IDM)~\cite{kesting2010enhanced}, a time-continuous car-following model which is widely used in traffic applications. We pick the IDM parameters such that the model is unstable below $18 \frac {\text m}{\text s}$, meaning that stop-and-go waves will propagate backward in the flux of traffic and grow instead of just dissipating. Numerous results demonstrate the string-unstable qualities of human-driver behavior, both via real human drivers in real life and via models such as IDM in simulation~\cite{cui2017stabilizing}.

\subsection{Lane-changing model}

We use a lane-changing model to enable more complex multi-lane dynamics for evaluation only. The model consists of a cut-in probability function $P_\text{in}(h, v_\text{lead})$ and a cut-out probability function $P_\text{out}(v_\text{lead})$, where $h$ is the space gap and $v_\text{lead}$ is the speed of the leading vehicle. Both are piecewise second-order polynomials whose coefficients were calibrated using data collected on the I-24 highway, which has not been published yet. At each time step $t$, and for each ego vehicle in the simulation, $P_\text{in}$ gives the probability that a vehicle cuts in front of the ego vehicle, while $P_\text{out}$ gives the probability that the leading vehicle cuts out. If a cut-in happens, a vehicle is inserted such that the ratio between the space gap of the inserted vehicle and the space gap of the ego vehicle after the cut-in follows a normal distribution (also fit to data), clipped to ensure safety after insertion. This model lets us measure the robustness of the control, allowing human vehicles to cut in front of the AV as it tries to open larger gaps to smooth out traffic waves.

\section{CONTROLLER DESIGN}

In this section, we formally define the problem in the context of reinforcement learning and discuss the structure and design of the controller.

\subsection{Defining the POMDP}

We use the standard RL formalism of maximizing the discounted sum of rewards for a finite-horizon partially-observed Markov decision process (POMDP).  We can formally define this POMDP as the tuple $(\mathcal{S},\mathcal{A} ,T ,R ,\gamma, \Omega ,\mathcal{O})$ where $\mathcal{S}$ is a set of states, $\mathcal{A}$ represents the set of actions, $T: \mathcal{S} \times \mathcal{A} \times \mathcal{S} \rightarrow \mathbb{R}$ represents the conditional probability distribution of transitioning to state $s'$ given state $s$ and action $a$, $R: S \times A \rightarrow \mathbb{R}$ is the reward function, and $\gamma \in (0, 1]$ is the discount factor used when calculating the sum of rewards. The final two terms are included since the state is hidden: $\Omega$ is the set of observations of the hidden state, and $\mathcal{O}: S \times \Omega \rightarrow \mathbb{R} $ represents the conditional observation probability distribution.

\subsubsection{Observation space}
The primary observation space (at time $t$) consists of the ego vehicle speed $v_t^\text{av}$, the leader vehicle speed $v_t^\text{lead}$, and the space gap (bumper-to-bumper distance) $h_t$ between the two vehicles. (Note that all distances are  in $\text m$, velocities in $\frac{\text m}{\text s}$, and accelerations in $\frac{\text m}{{\text s}^2}$.) Two gap thresholds are also included: $h_t^\text{min}$, which is the failsafe threshold below which the vehicle will always brake, and $h_t^\text{max}$, the gap-closing threshold above which the vehicle will always accelerate. We also include the history of the ego vehicle's speed over the last 0.5 seconds. Finally, the observation space includes traffic information from the speed planner. This consists of the current target speed $v_t^\text{sp}$, as well as the target speeds 200m, 500m, and 1km downstream of the vehicle's current position. Note that the AV only observes its leading vehicle and that there is no explicit communication between AVs. All observations provided to the RL agent are rescaled to the range $[-1,1]$.

\subsubsection{Action space}
\label{sec:action_space}
The action space consists of an instantaneous acceleration $a_t \in [-3, 1.5]$. After the RL output, gap-closing and failsafe wrappers are then enforced. We define the gap-closing wrapper such that $h_t^\text{max} = \displaystyle \max(120, 6 v_t^\text{av})$, meaning that the AV will be forced to accelerate if its space gap becomes larger than $120$m or its time gap larger than $6$s. This result is then wrapped within a failsafe, which enforces safe following distances and prevents collisions. We define the time to collision $\Delta^\text{TTC}_t$ of the ego and lead vehicles as:
$$\Delta^\text{TTC}_t = \begin{cases} 
\displaystyle \frac{h_t}{v_t^\text{diff}} & \text{if } v_t^\text{diff} > 0 \\
+\infty & \text{otherwise}
\end{cases}$$
$$v_t^\text{diff} = \left[v_t^\text{av}\left(1 + \frac{4}{30}\right) + 1\right] - v_t^\text{lead}$$
where the AV velocity is slightly exaggerated to ensure robustness at both low and high speeds. The actual numbers are chosen heuristically: for instance, if both AV and leader drive at 30$\frac{\text{m}}{\text{s}}$, the failsafe will ensure a minimum gap of 30$\text{m}$. 
The failsafe triggers if the time to collision ever goes below 6 seconds, in which case the RL output is overridden and the vehicle will brake at its maximum allowed deceleration.
We thus have $h_t^\text{min} = 6 v_t^\text{diff}$. 
The final RL acceleration is given by:
$$a_t^\text{out} =  \begin{cases} 
      -3 & \text{if } \Delta^\text{TTC}_t \leq 6 \quad (\Leftrightarrow h_t \leq h_t^\text{min} ) \\
      1.5 & \text{if } \Delta^\text{TTC}_t > 6 \text{ and } h_t \geq h_t^\text{max} \\
      a_t & \text{otherwise}  \\
   \end{cases}
 $$
 which is further clipped to ensure that the speed $v^\text{av}_t$ remains within the boundaries $[0, 35]\frac{\text{m}}{\text{s}}$. Note that the free-flow behavior due to the gap-closing wrapper will be to drive at the speed limit in the absence of a leader.

\subsubsection{Optimization criterion}

At the core, we aim to minimize the overall energy consumption of the traffic. However, as sparse rewards are harder to optimize, we employ proxies that can be minimized at each time step. We mainly aim to minimize the instantaneous energy consumption of the AV and a platoon of vehicles behind it. For comfort, and as another proxy for energy savings, we also minimize squared acceleration amplitudes. Since optimizing for energy and acceleration can be done by stopping or maintaining unreasonably large or small gaps for comfort, we penalize gaps outside of a certain range; this also penalizes the use of failsafe and gap-closing interventions. To further discourage large gaps within this allowed range, the final term adds a penalty proportional to the time gap (space gap divided by ego speed). This is formalized as the reward function $r_t$, which is given by:
\begin{align*}
    \displaystyle r_t = & - c_1 \frac{1}n \sum_{i=1}^{n} E_t^i - c_2 (a_t^\text{out})^2 - c_3 \mathds{1} \left[{h_t \notin [h_t^\text{min}, h_t^\text{max}]} \right] \\ & - c_4\frac{h_t}{v_t^\text{av}}\mathds{1} \left[h_t > 10 \land v_t^\text{av} > 1\right]
\end{align*}

where $E_t^i$ is the instantaneous energy consumption of vehicle $i$ at time $t$, where index $i=1$ corresponds to the AV, and indexes $i=2$ to $i=n$ correspond to the following $n-1$ IDM vehicles. $\mathds{1}$ is defined such that $\mathds{1} \left[\mathcal{P}\right] = 1 \text{ if } \mathcal P \text{ is true, } 0 \text{ otherwise}$.

\subsection{Training algorithm}

We use single-agent Proximal Policy Optimization~\cite{schulman2017ppo} (PPO) with an augmented value function as our training algorithm. PPO is a policy gradient algorithm, a class of RL techniques that optimize for cumulative, discounted reward via shifting the parameters of the neural net directly. More explicitly, policy gradient methods represent the policy as $\pi_\theta(a | s)$, where $\theta$ are the parameters of the neural net.

\begin{figure*} [h]
    \centering
    \includegraphics[width=\linewidth]{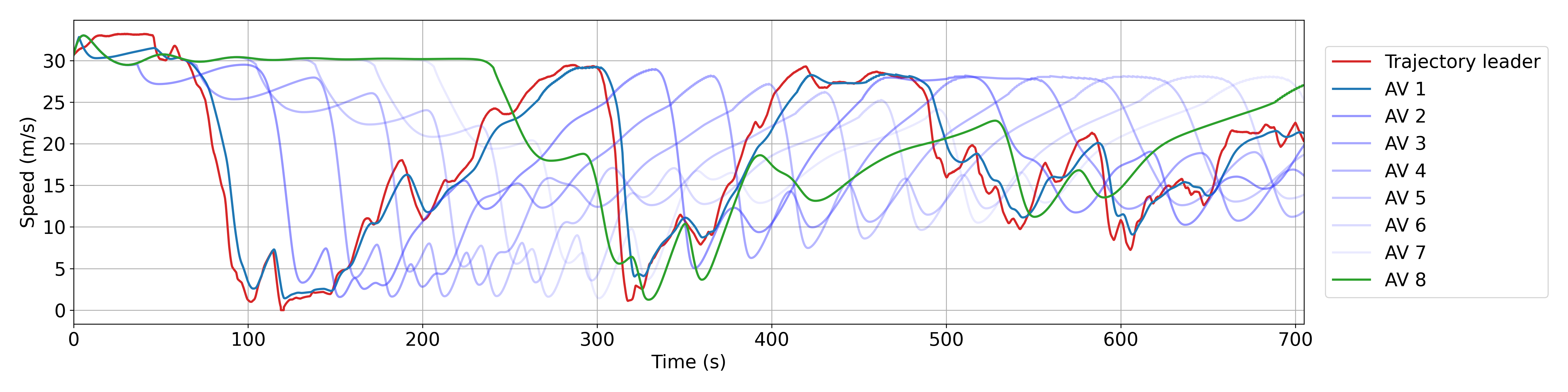}
    \caption{Evolution of the speed of the trajectory leader and the AVs in a platoon of 200 vehicles. The 8 AVs are equally spaced at a 4\% penetration rate. The first AV in the platoon is shown in blue, the following ones are displayed by decreasing opacity and the last one is in green, demonstrating the smoothing effect of the AVs on the leader trajectory. In particular, one can see how the first AV (in blue) already smoothes the trajectory leader (in red), doesn't slow down as much or accelerate as fast, and thus saves energy.}
    \label{fig:av_speeds}
\end{figure*}

\begin{figure}
    \centering
    \includegraphics[width=1.0\linewidth]{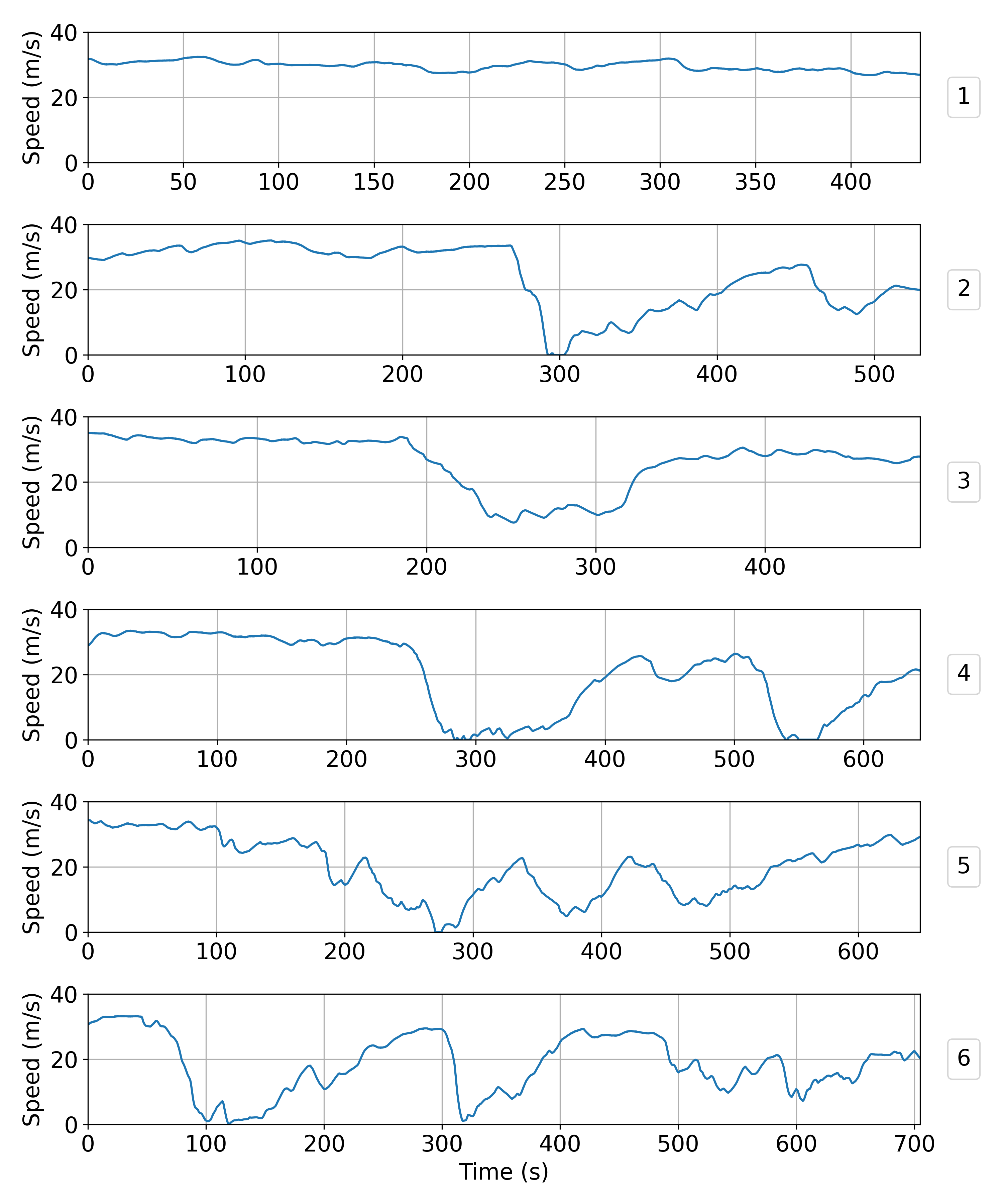}
    \caption{Trajectories used for evaluation, numbered from 1 to 6. The first one corresponds to free flow, while the five others contain both low and high speeds, including sharp breaking, sharp accelerating or stop-and-go behaviors.}
    \label{fig:eval_trajs}
\end{figure}

During training, we give additional observations that are available in simulation as an input to the value network. This includes the cumulative miles traveled and gallons of gas consumed, which allows for estimating energy efficiency. This augmented observation space also includes the following metadata: the size of the finite horizon, an identifier for the specific trajectory chunk being trained on, and the vehicle's progress (in space and time) within this chunk. The additional information removes some of the partial observability from the system, allowing the value function to more accurately predict the factors that influence reward. 

\subsection{Experiment details}

We run the PPO experiments using the implementation provided in Stable Baselines 3\footnote{\url{https://github.com/DLR-RM/stable-baselines3}} version 1.6.2. We train the model for 2500 iterations, which takes about 6 hours on 90 CPUs. We use a training batch size of 9000, a batch size of 3000, a learning rate of $3 \cdot 10^{-4}$ and do $5$ epochs per iteration. The simulation horizon is set to $500$, and we do $10$ simulation steps per environment step, ie. each action is repeated $10$ times. Given that the simulation time step is $dt=0.1$s, this means that the action changes only every second during training. This allows us to artificially reduce the horizon so that it only is $50$, meaning that each training batch contains 100 simulation rollouts. The agent's policy is modeled as a fully-connected neural network with 4 hidden layers of 64 neurons each and $\tanh$ linearities, with continuous actions. The value network has the same architecture as the policy network. We set the discount factor $\gamma$ to $0.999$, the GAE value $\lambda$ to $0.99$, and the other training and PPO parameters are left to their default values. 

We train with a platoon of $n=25$ vehicles (not including the leading trajectory vehicle) consisting of one AV followed by 24 IDM vehicles. 
For our reward function, we used coefficients $c_1 = 0.06$, $c_2 = 0.02$, $c_3 = 0.6$, $c_4 = 0.005$. Both the model parameters and training hyperparameters are determined through grid search, with each experiment conducted using 4 to 8 distinct random seeds to overcome instances of the agent getting trapped in local optima.

\begin{figure*}[htp]
  \centering
  \subfigure{\includegraphics[width=0.45\linewidth]{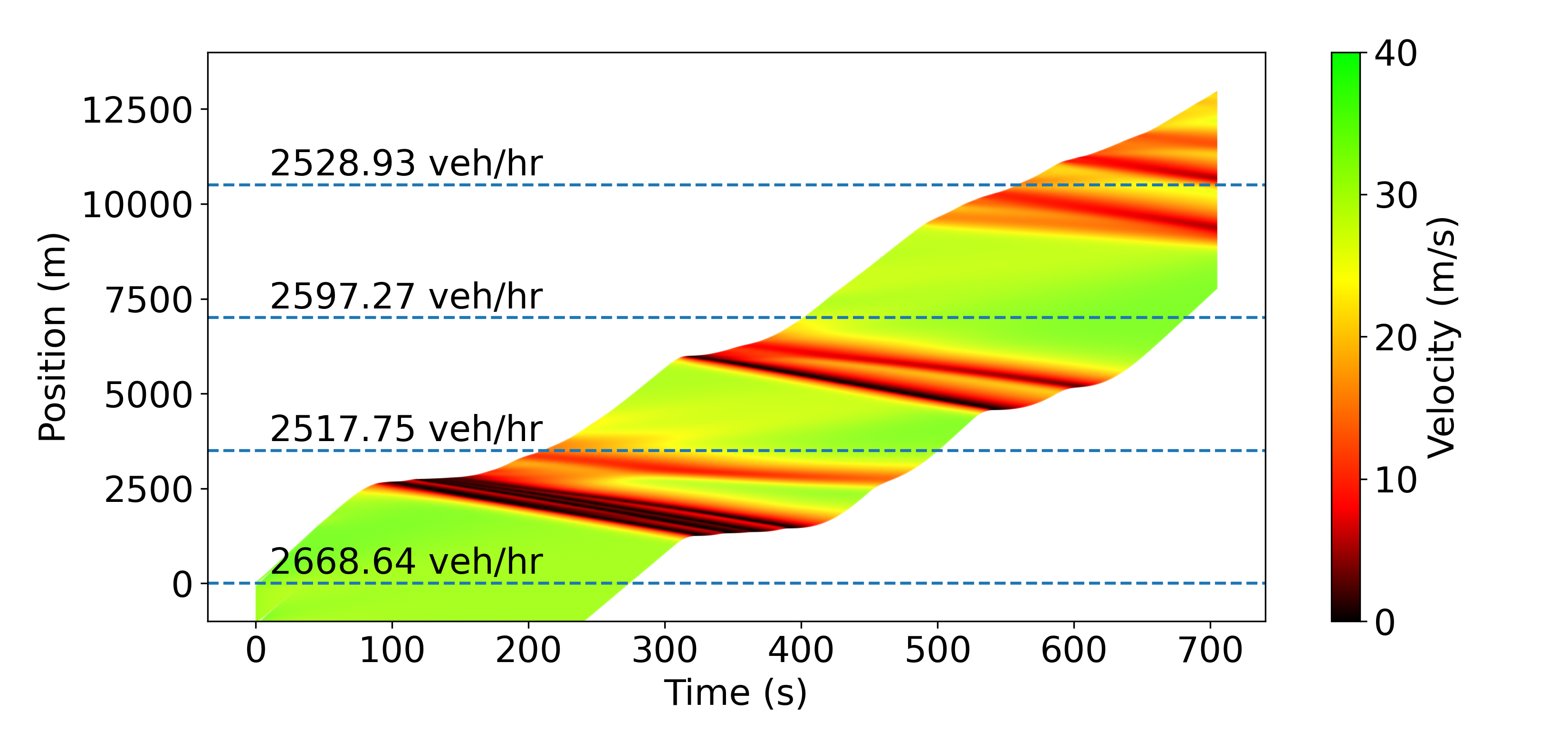}}
  \subfigure{\includegraphics[width=0.45\linewidth]{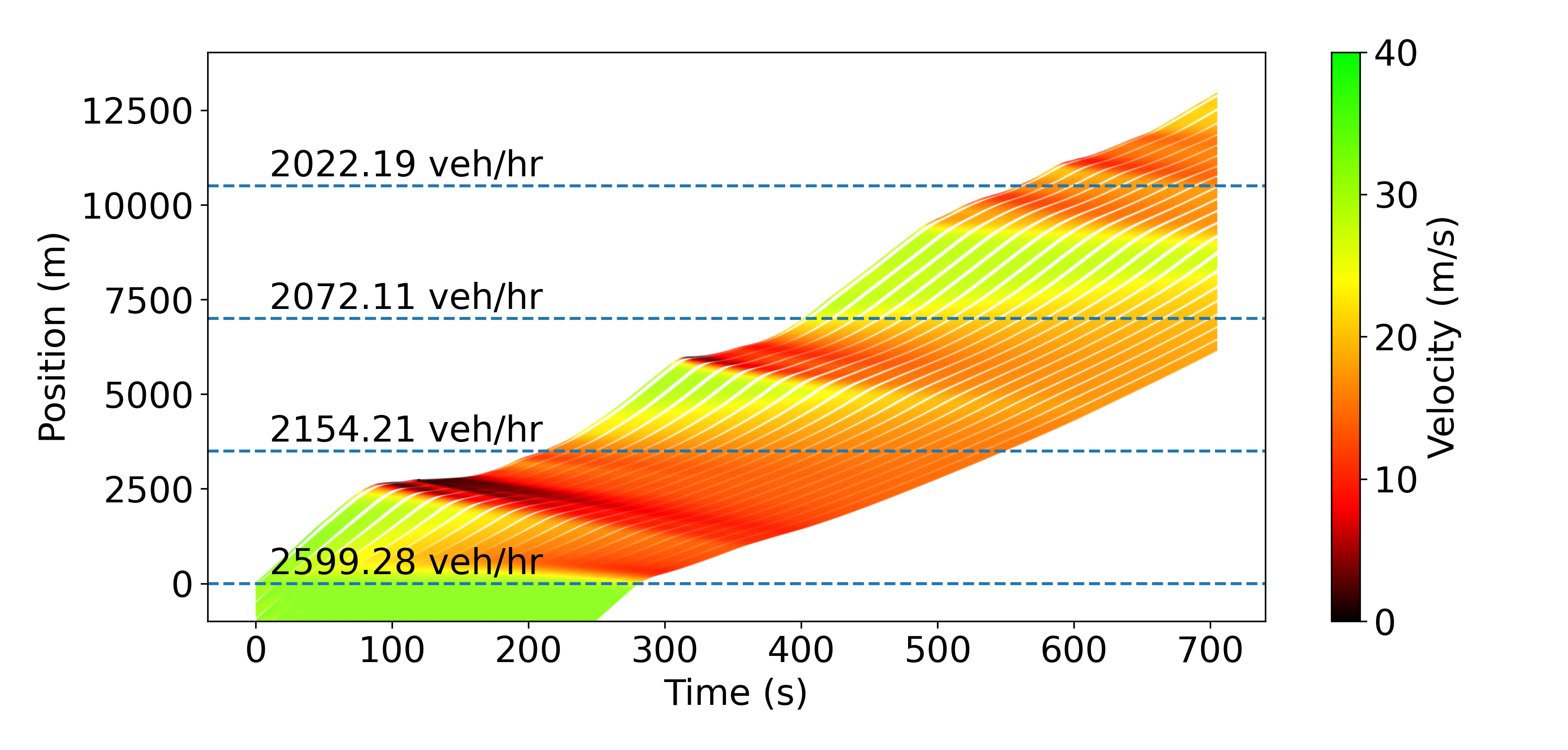}}
  \subfigure{\includegraphics[width=0.45\linewidth]{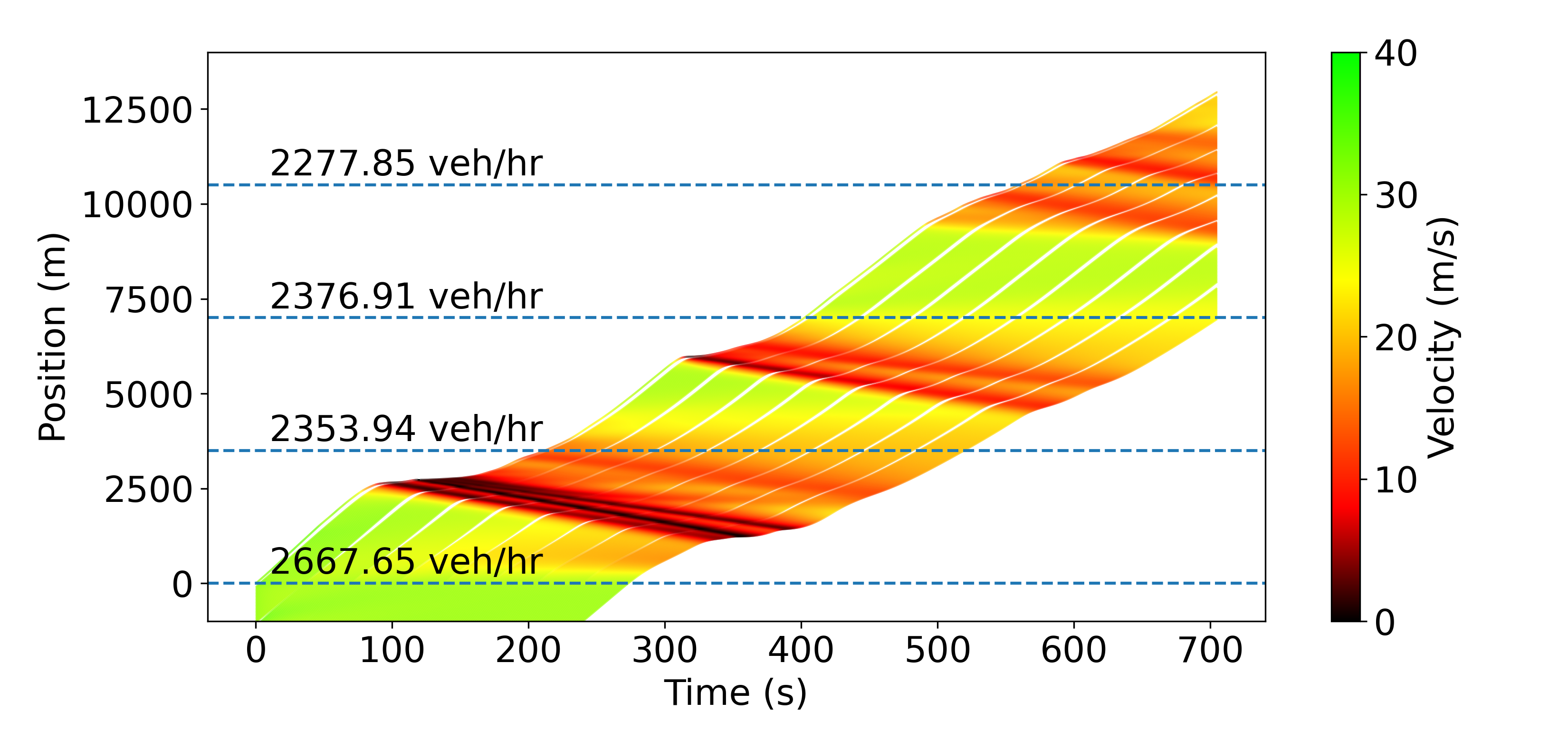}}
  \subfigure{\includegraphics[width=0.45\linewidth]{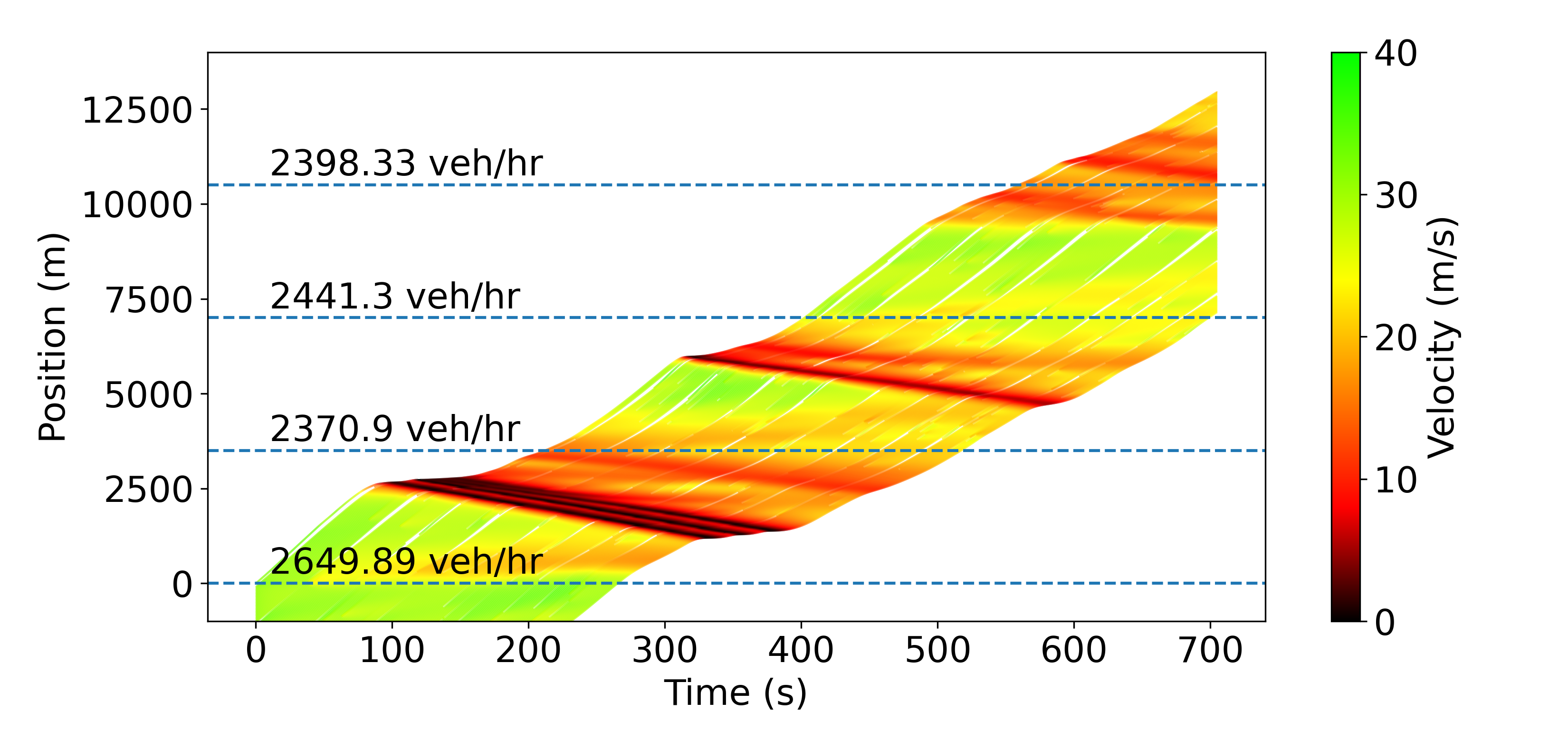}}
  \caption{Time-space diagrams, each representing a simulation of 200 vehicles. Each vehicle trajectory is plotted as a line in position-by-time space, with color representing the speed of that vehicle. One can observe the wave-smoothing effect of the RL-controlled AVs over time. Horizontal lines display the throughput of the traffic flow at that particular position. Also note that as a warm-up, all AVs behave as humans when their position is negative. \textbf{Top left: } all 200 vehicles are IDMs. Traffic waves are illustrated by the red and black colors, while bright green represents free flow. \textbf{Top right: } 20 equally-spaced RL-controlled AVs (10\% penetration rate, 1 AV every 10 vehicles). One can see the larger gaps opened by the AVs as the white lines between platoons. \textbf{Bottom left: } 8 equally-spaced RL-controlled AVs (4\% penetration rate, 1 AV every 25 vehicles). \textbf{Bottom right: } 8 equally-spaced RL-controlled AVs (4\% penetration rate) with the lane-changing model enabled (note that it is disabled in all 3 other subfigures).}
  \label{fig:tsd}
\end{figure*}

\section{RESULTS}

In this section, we analyze the performances of our RL controller in simulation, in terms of energy savings, wave-smoothing, behavior and robustness. 

The smoothing effect of the AVs is illustrated in Fig.~\ref{fig:av_speeds}, where one can see the speeds of all the AVs in a platoon of 200 vehicles as a function of time on trajectory 6 (see Fig.~\ref{fig:eval_trajs}). One can observe how the speed profiles become smoother and smoother after each AV. 

Fig.~\ref{fig:tsd} illustrates the smoothing performed by the RL agents on trajectory 6 in a different way. The top-left time-space diagram shows that the humans don't smooth any waves; on the contrary, they even create some due to the string-unstable nature of the IDM we use. At a 10\% penetration rate, most of the waves get smoothed out, less at 4\%, and even less with lane-changing. However, in all 3 cases, the diagrams clearly demonstrate the improvement over the baseline. As expected, AVs opening larger gaps also leads to decreased throughput, and the best throughput is achieved when the lane-changing model is enabled and human vehicles fill in the gaps. This comes down to a trade-off between throughput reduction and energy savings, which can be tuned by varying the penetration rate. 

Table~\ref{table:mpg} shows the energy savings that our controller achieves on the evaluation trajectories (shown in Fig.~\ref{fig:eval_trajs}) when deployed on AVs at two different penetration rates, with and without lane-changing enabled. The percentages correspond to how much the average system miles-per-gallon (MPG) value increases when AVs use our RL controller, compared to when they all behave as humans. The average MPG is defined as the sum of the distances traveled by all the vehicles in the simulation, divided by the sum of their energy consumptions. 

\begin{table}
\centering
\begin{tabular}{ |c||c|c||c|c| } 
 \hline
 Index & 10\% w/o LC & 10\% w/ LC & 4\% w/o LC & 4\% w/ LC \\  \hline 
 1 & +7.33\% & +8.39\% & +4.29\% & +6.12\% \\  \hline
 2 & +10.87\% & +12.63\% & +7.04\% & +9.42\% \\  \hline
 3 & +14.65\% & +14.48\% & +9.02\% & +7.23\% \\  \hline
 4 & +15.02\% & +13.19\% & +9.23\% & +8.54\% \\  \hline
 5 & +22.58\% & +15.77\% & +17.05\% & +8.37\% \\  \hline
 6 & +28.98\% & +18.55\% & +19.96\% & +15.40\% \\  \hline
\end{tabular}
\caption{\label{table:mpg} We run simulations with 200 vehicles, and show the improvement in system MPG when we control 10\% (Left) or 4\% (Right) of the vehicles, compared to when all vehicles behave as IDMs, with and without lane-changing in both cases. The trajectories used for evaluation can be seen in Fig.~\ref{fig:eval_trajs}, with corresponding indexes.}
\end{table}

We can observe the energy savings varying a lot depending on the trajectories, which is expected since trajectories that are mostly free flow (like trajectory 1) cannot be improved much, while one with a lot of stop-and-go waves (like trajectory 6) has a lot of potential to be smoothed. As one can expect, energy savings decrease as the AV penetration rate decreases or as lane-changing is enabled, but even at 4\% penetration and with lane-changing, the controller reduces the energy consumption on trajectory 6 by over 15\%, while only reducing throughput by 5\%. 

We also note that the controller appears robust to not having access to the speed planner. For example at a 4\% penetration rate and without lane-changing, the control achieves +16.87\% energy improvement on trajectory 6 without the speed planner (compared to +19.96\% with the speed planner), and the trend is similar on the other trajectories.

\begin{figure}
    \centering
    \includegraphics[width=\linewidth]{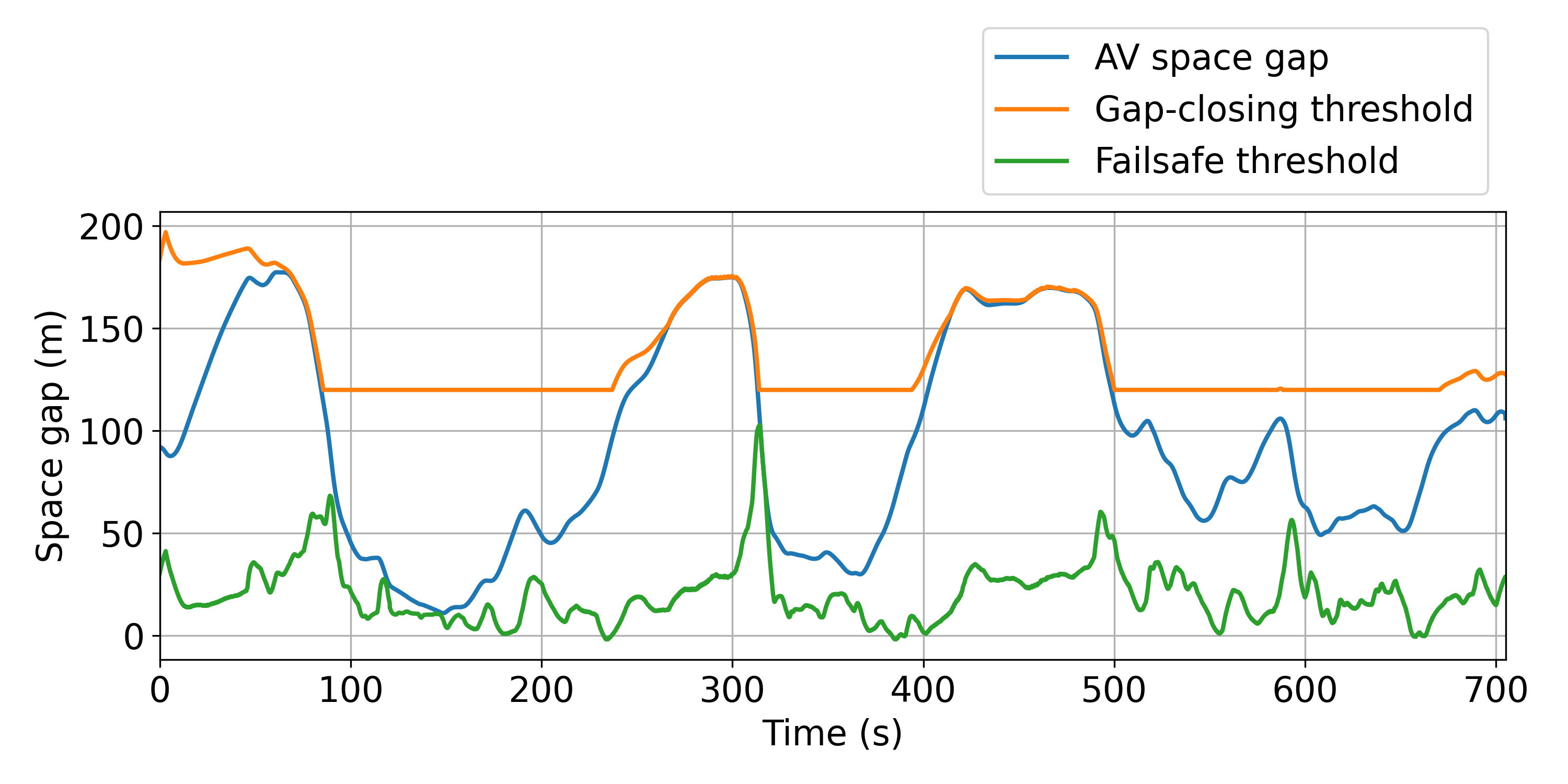}
    \caption{Space gap of the first AV in the platoon (in blue) by time, on the same scenario as in Fig.~\ref{fig:av_speeds}. The orange line shows the gap-closing threshold $h_t^\text{max}$ and the green line shows the failsafe threshold $h_t^\text{min}$, introduced in Sec.~\ref{sec:action_space}.}
    \label{fig:av_gap}
\end{figure}

Finally, in Fig.~\ref{fig:av_gap}, we illustrate the gaps opened by the AV on trajectory 6, along with the failsafe and gap-closing thresholds. The gap-closing threshold allows the AV to open larger gaps and consequently absorb abrupt braking from its leader while ensuring that these gaps are not overly large. As can be expected, we have observed that the larger the maximum gap we allow, the better the AV performs in terms of energy savings. However, a larger maximum gap is usually accompanied by a decrease in throughput, which is again a trade-off. The failsafe threshold mostly ensures safety and comfort for the driver, although it is worth noting that when deploying the controller, we integrate an additional explicit safety wrapper, as detailed in~\cite{lichtle2022deploying}. 

\section{CONCLUSION}

In this work, we developed RL policies that incorporate both accessible local observations and downstream traffic information, achieving substantial energy savings in simulation. There are several avenues for future research. Given that all the training is conducted in a simulated environment, it would be beneficial to train the agent with more realistic dynamics by enhancing the accuracy of the various models that make up this simulation, like human driving behavior, lane-changing dynamics, and energy consumption metrics. Additionally, it would be interesting to explore multi-agent RL to help the model be robust to interactions between AVs and potentially enable cooperation between them. 

While our simulation process has a distinct speed advantage over large micro-simulators, it could benefit significantly from vectorization. Moreover, despite our technical capacity to deploy our controller safely onto a real-world vehicle, gathering results is challenging and necessitates further field tests.
Another direction of research we are exploring consists of training and deploying adaptive cruise control (ACC)-based controllers, where the policy outputs a desired set-speed instead of an acceleration. By design of the ACC, the control would be safe and smooth, and easily deployable 
at a large scale simply by augmenting the onboard ACC algorithm to use the RL control as a set-speed actuator. 

\bibliographystyle{IEEEtran}
\bibliography{references.bib}

\begin{thebibliography}{10}
\providecommand{\url}[1]{#1}
\csname url@samestyle\endcsname
\providecommand{\newblock}{\relax}
\providecommand{\bibinfo}[2]{#2}
\providecommand{\BIBentrySTDinterwordspacing}{\spaceskip=0pt\relax}
\providecommand{\BIBentryALTinterwordstretchfactor}{4}
\providecommand{\BIBentryALTinterwordspacing}{\spaceskip=\fontdimen2\font plus
\BIBentryALTinterwordstretchfactor\fontdimen3\font minus
  \fontdimen4\font\relax}
\providecommand{\BIBforeignlanguage}[2]{{%
\expandafter\ifx\csname l@#1\endcsname\relax
\typeout{** WARNING: IEEEtran.bst: No hyphenation pattern has been}%
\typeout{** loaded for the language `#1'. Using the pattern for}%
\typeout{** the default language instead.}%
\else
\language=\csname l@#1\endcsname
\fi
#2}}
\providecommand{\BIBdecl}{\relax}
\BIBdecl

\bibitem{eiaEnergyTransportation}
``Use of energy for transportation - u.s. energy information administration
  (eia) --- eia.gov,''
  \url{https://www.eia.gov/energyexplained/use-of-energy/transportation.php},
  [Accessed 04-Mar-2023].

\bibitem{eiaUSEnergy}
``{U}.{S}. energy facts explained - consumption and production - {U}.{S}.
  {E}nergy {I}nformation {A}dministration ({E}{I}{A}) --- eia.gov,''
  \url{https://www.eia.gov/energyexplained/us-energy-facts/}, [Accessed
  04-Mar-2023].

\bibitem{epaHighlightsAutomotive}
``{H}ighlights of the {A}utomotive {T}rends {R}eport | {U}{S} {E}{P}{A} ---
  epa.gov,''
  \url{https://www.epa.gov/automotive-trends/highlights-automotive-trends-report},
  [Accessed 04-Mar-2023].

\bibitem{naicAutonomousVehicles}
``{A}utonomous {V}ehicles --- content.naic.org,''
  \url{https://content.naic.org/cipr-topics/autonomous-vehicles}, [Accessed
  04-Mar-2023].

\bibitem{beaty1998traffic}
W.~Beaty, ``Traffic “experiments” and a cure for waves \& jams,''
  \url{http://amasci.com/amateur/traffic/trafexp.html}, 1998, [Accessed
  15-Oct-2006].

\bibitem{wu2019tracking}
F.~Wu, R.~E. Stern, S.~Cui, M.~L. Delle~Monache, R.~Bhadani, M.~Bunting,
  M.~Churchill, N.~Hamilton, B.~Piccoli, B.~Seibold \emph{et~al.}, ``Tracking
  vehicle trajectories and fuel rates in phantom traffic jams: Methodology and
  data,'' \emph{Transportation Research Part C: Emerging Technologies},
  vol.~99, pp. 82--109, 2019.

\bibitem{Jang2018SimulationVehicles}
K.~Jang, E.~Vinitsky, B.~Chalaki, B.~Remer, L.~Beaver, A.~Malikopoulos, and
  A.~Bayen, ``{Simulation to scaled city: zero-shot policy transfer for traffic
  control via autonomous vehicles},'' in \emph{2019 International Conference on
  Cyber-Physical Systems}, Montreal, CA, 2018.

\bibitem{vinitsky2018lagrangian}
E.~Vinitsky, K.~Parvate, A.~Kreidieh, C.~Wu, and A.~Bayen, ``Lagrangian control
  through deep-rl: Applications to bottleneck decongestion,'' in \emph{2018
  21st International Conference on Intelligent Transportation Systems
  (ITSC)}.\hskip 1em plus 0.5em minus 0.4em\relax IEEE, 2018, pp. 759--765.

\bibitem{stern2018dissipation}
R.~E. Stern, S.~Cui, M.~L. Delle~Monache, R.~Bhadani, M.~Bunting, M.~Churchill,
  N.~Hamilton, H.~Pohlmann, F.~Wu, B.~Piccoli \emph{et~al.}, ``Dissipation of
  stop-and-go waves via control of autonomous vehicles: Field experiments,''
  \emph{Transportation Research Part C: Emerging Technologies}, vol.~89, pp.
  205--221, 2018.

\bibitem{cui2017stabilizing}
S.~Cui, B.~Seibold, R.~Stern, and D.~B. Work, ``Stabilizing traffic flow via a
  single autonomous vehicle: Possibilities and limitations,'' in
  \emph{Intelligent Vehicles Symposium (IV), 2017 IEEE}.\hskip 1em plus 0.5em
  minus 0.4em\relax IEEE, 2017, pp. 1336--1341.

\bibitem{gu2017deep}
S.~Gu, E.~Holly, T.~Lillicrap, and S.~Levine, ``Deep reinforcement learning for
  robotic manipulation with asynchronous off-policy updates,'' in
  \emph{Robotics and Automation (ICRA), 2017 IEEE International Conference
  on}.\hskip 1em plus 0.5em minus 0.4em\relax IEEE, 2017, pp. 3389--3396.

\bibitem{vinyals2019grandmaster}
O.~Vinyals, I.~Babuschkin, W.~M. Czarnecki, M.~Mathieu, A.~Dudzik, J.~Chung,
  D.~H. Choi, R.~Powell, T.~Ewalds, P.~Georgiev \emph{et~al.}, ``Grandmaster
  level in starcraft ii using multi-agent reinforcement learning,''
  \emph{Nature}, vol. 575, no. 7782, pp. 350--354, 2019.

\bibitem{silver2017mastering}
D.~Silver, J.~Schrittwieser, K.~Simonyan, I.~Antonoglou, A.~Huang, A.~Guez,
  T.~Hubert, L.~Baker \emph{et~al.}, ``Mastering the game of go without human
  knowledge,'' \emph{Nature}, vol. 550, no. 7676, p. 354, 2017.

\bibitem{wu2017flow}
C.~Wu, A.~Kreidieh, K.~Parvate, E.~Vinitsky, and A.~M. Bayen, ``Flow:
  Architecture and benchmarking for reinforcement learning in traffic
  control,'' \emph{arXiv preprint arXiv:1710.05465}, p.~10, 2017.

\bibitem{vinitsky2018benchmarks}
E.~Vinitsky, A.~Kreidieh, L.~Le~Flem, N.~Kheterpal, K.~Jang, C.~Wu, F.~Wu,
  R.~Liaw, E.~Liang, and A.~M. Bayen, ``Benchmarks for reinforcement learning
  in mixed-autonomy traffic,'' in \emph{Conference on Robot Learning}.\hskip
  1em plus 0.5em minus 0.4em\relax PMLR, 2018, pp. 399--409.

\bibitem{sugiyama2008traffic}
Y.~Sugiyama, M.~Fukui, M.~Kikuchi, K.~Hasebe, A.~Nakayama, K.~Nishinari, S.-i.
  Tadaki, and S.~Yukawa, ``Traffic jams without bottlenecks—experimental
  evidence for the physical mechanism of the formation of a jam,'' \emph{New
  journal of physics}, vol.~10, no.~3, p. 033001, 2008.

\bibitem{lichtle2022deploying}
N.~Lichtl{\'e}, E.~Vinitsky, M.~Nice, B.~Seibold, D.~Work, and A.~M. Bayen,
  ``Deploying traffic smoothing cruise controllers learned from trajectory
  data,'' in \emph{2022 International Conference on Robotics and Automation
  (ICRA)}.\hskip 1em plus 0.5em minus 0.4em\relax IEEE, 2022, pp. 2884--2890.

\bibitem{matthew_nice_2021_6456348}
\BIBentryALTinterwordspacing
M.~Nice, N.~Lichtlé, G.~Gumm, M.~Roman, E.~Vinitsky, S.~Elmadani, M.~Bunting,
  R.~Bhadani, K.~Jang, G.~Gunter \emph{et~al.}, ``The i-24 trajectory
  dataset,'' Sep. 2021. [Online]. Available:
  \url{https://doi.org/10.5281/zenodo.6456348}
\BIBentrySTDinterwordspacing

\bibitem{fu2023planner}
Z.~Fu, A.~R. Kreidieh, H.~Wang, J.~W. Lee, M.~L.~D. Monache, and A.~M. Bayen,
  ``Cooperative driving for speed harmonization in mixed-traffic
  environments,'' 2023.

\bibitem{lee2021energy}
J.~W. Lee, G.~Gunter, R.~Ramadan, S.~Almatrudi, P.~Arnold, J.~Aquino,
  W.~Barbour, R.~Bhadani, J.~Carpio, F.-C. Chou \emph{et~al.}, ``Integrated
  framework of vehicle dynamics, instabilities, energy models, and sparse flow
  smoothing controllers,'' in \emph{Proceedings of the Workshop on Data-Driven
  and Intelligent Cyber-Physical Systems}, 2021, p. 41–47.

\bibitem{kesting2010enhanced}
A.~Kesting, M.~Treiber, and D.~Helbing, ``Enhanced intelligent driver model to
  access the impact of driving strategies on traffic capacity,''
  \emph{Philosophical Transactions of the Royal Society A: Mathematical,
  Physical and Engineering Sciences}, vol. 368, no. 1928, pp. 4585--4605, 2010.

\bibitem{schulman2017ppo}
J.~Schulman, F.~Wolski, P.~Dhariwal, A.~Radford, and O.~Klimov, ``Proximal
  policy optimization algorithms,'' \emph{arXiv preprint arXiv:1707.06347},
  2017.

\end{thebibliography}

\end{document}